\documentclass[conference]{IEEEtran}

\usepackage[backend=bibtex,
     bibstyle=ieee,
     sortcites=true,
     maxnames=2,
     maxbibnames=12,
     minnames=1,
     mincrossrefs=99,
     uniquelist=false,
     defernumbers=true
]{biblatex}

\makeatletter
\newcounter{IEEE@bibentries}
\renewcommand\IEEEtriggeratref[1]{%
 \renewbibmacro{finentry}{%
   \stepcounter{IEEE@bibentries}%
   \ifthenelse{\equal{\value{IEEE@bibentries}}{#1}}
   {\finentry\@IEEEtriggercmd}
   {\finentry}%
 }%
}
\makeatother

\usepackage{amsmath,amssymb,amsfonts}
\usepackage{booktabs}
\usepackage{tabularx}
\usepackage{csquotes}

\usepackage{tikz}
\usepackage{tikzpeople}
\usetikzlibrary{trees}
\usetikzlibrary{positioning}
\usetikzlibrary{arrows.meta,arrows}
\usetikzlibrary{decorations.pathreplacing}
\usepackage[edges]{forest}
\forestset{qtree/.style={for tree={parent anchor=south,
           child anchor=north,align=center,inner sep=0pt}}}

\usepackage{pgfplots}
\pgfplotsset{compat=1.18}
\usetikzlibrary{
        pgfplots.dateplot,
        pgfplots.groupplots,
        pgfplots.statistics,
        external,
    }

\usepackage{pgfplotstable}

\usepackage{cleveref}
\crefname{figure}{\figurename}{figures}
\Crefname{figure}{\figurename}{Figures}
\crefname{table}{\tablename}{Tables}

\usepackage{xspace}

\newcommand*{\ie}{i.e.\@\xspace}

\usepackage{subcaption}

\definecolor{sron0}{HTML}{332288}
\definecolor{sron1}{HTML}{88CCEE}
\definecolor{sron2}{HTML}{117733}
\definecolor{sron3}{HTML}{DDCC77}
\definecolor{sron4}{HTML}{CC6677}
\definecolor{sron5}{HTML}{AA4499}

\begin{document}

\author{\IEEEauthorblockN{Manuel Wedler}
\IEEEauthorblockA{\textit{Humboldt-University of Berlin}\\
manuel@wedler.dev}
\and
\IEEEauthorblockN{Erik Daniel}
\IEEEauthorblockA{
\textit{TU Dresden}\\
erik.daniel@tu-dresden.de}
\and
\IEEEauthorblockN{Florian Tschorsch}
\IEEEauthorblockA{\textit{TU Dresden}\\
florian.tschorsch@tu-dresden.de}
}

\title{Plausibly Deniable Content Discovery for Bitswap Using Random Walks}

\maketitle

\begin{abstract}
Bitswap is the data exchange protocol for the content-addressed peer-to-peer overlay network IPFS.
During content discovery, Bitswap reveals the interest of a peer in content to all neighbors, enabling the tracking of user interests.
In our paper, we propose a modification of the Bitswap protocol, which enables source obfuscation using proxies for content discovery.
The proxies are selected via a random-walk.
Enabling content discovery through proxies introduces plausible deniability.
We evaluate the protocol modification with a simulation.
The protocol modification demonstrates enhanced privacy, while maintaining acceptable performance levels.
\end{abstract}


\section{Introduction}
Peer-to-peer~(P2P) overlay networks can be used for decentralized data exchange, providing an alternative to centralized data storage.
Due to the P2P architecture, anyone can join the network to exchange data.
A prime example for such a network with many active peers is the InterPlanetary File System~(IPFS)~\cite{benet2014ipfs}.
IPFS incorporates content-addressed exchange of data, using the Bitswap protocol for content retrieval.

Bitswap operates on blocks, which are data chunks each identifiable by a self-verifiable content identifier~(CID).
The CID is used for content discovery and retrieval.
For content discovery, Bitswap queries all neighbors for a CID.
If this initial request yields no providers, Bitswap asks the Kademlia-based Distributed Hash Table~(DHT) of IPFS for content providers.
While sending the request to all neighbors improves fault tolerance and speed of the content discovery process~\cite{henningsen2020mapping,trautwein2022design}, it also reveals the interest to many unconcerned peers.
It is possible for passive participants to observe large parts of request~\cite{balduf2021monitoring}, enabling profiling of peers.

In this paper, our goal is to improve the privacy of content discovery.
We therefore introduce \emph{RaWa-Bitswap}, a random walk-based proxy approach.
We propose to outsource content discovery to other peers, proxies, which perform the lookup on behalf of the peer.
The proxy is chosen via a random walk, relaying the request over peers to the proxy.
This obfuscates the original source of a request and provides each involved peer with plausible deniability.
Any peer, requestor, relay, or proxy, can plausibly claim that the lookup itself is executed on behalf of another peer.
Such a random walk is also used by Dandelion~\cite{bojja2017dandelion} and Dandelion++~\cite{fanti2018dandelion++} to improve the transaction privacy in a cryptocurrency.
In Dandelion/Dandelion++, transactions are relayed through the P2P network via a random walk to a proxy which diffuses the transaction.

In contrast to Dandelion, we use a random walk to select a proxy for content discovery to improve the privacy of Bitswap.
More specifically, RaWa-Bitswap forwards requests on a random walk to a proxy.
The proxy performs the normal Bitswap content discovery and returns information about content providers over the same path back to the requestor.
The requester can use the information to directly retrieve content, revealing the peer's interest to only one content provider. 

We evaluate RaWa-Bitswap with a simulation and compare RaWa-Bitswap to the normal Bitswap in terms of privacy and performance properties.
RaWa-Bitswap significantly enhances privacy for Bitswap users against passive adversaries, while preserving performance under the tested conditions.
Therefore, RaWa-Bitswap shows reduced detection probabilities of individual peer's interests and an increased level of plausible deniability.
Due to the limited forwarding of the request, the load on the network imposed by the method is significantly lower than similar network-level privacy-enhancements~\cite{de2021accelerating,daniel2023improving}.

As our contribution, we propose a design of a random walk-based forwarding protocol, enhancing privacy under a request-response model.
We provide a proof-of-concept~(PoC) implementation of the proposed design based on Bitswap of the boxo library~\cite{gitgoboxo} v0.8.0.
Furthermore, we evaluate the effect of the changed content discovery on privacy and performance.

The remainder is structured as follows:
In \Cref{sec:relwork}, we present related work.
\Cref{sec:method} explains the functionality of RaWa-Bitswap and \Cref{sec:impl} explains details of the PoC.
In \Cref{sec:eval}, we evaluate our approach using a simulation.
\Cref{sec:conclusion} concludes the paper.

\section{Related Work}\label{sec:relwork}
The selection of a random node of the network to handle requests to obfuscate the source of the request is a common method.
In combination with layer encryption, this can lead to unlinkability
as in Tor~\cite{dingledine2004tor}.
A similar idea is used by Tarzan~\cite{freedman2002tarzan}, ShadowWalker~\cite{mittal2009shadowwalker}, and Torsk~\cite{mclachlan2009scalable} where nodes are randomly selected from the P2P network.
Although, the selection process of Torsk is vulnerable to denial-of-service attacks~\cite{wang2010search}.
All these methods and proposals use rather a random node selection, where the selection process is completely controlled by the origin, than a random walk.
Through such methods, the origin cannot deny that it executes a request.

A random walk, where each node chooses the next hop by itself provides plausible deniability for all peers on the path.
An early example is Crowds~\cite{reiter1998crowds}, where requests are forwarded over a random walk to a proxy, which executes a web transaction on behalf of another user.
The next hop of the random walk is chosen from all known peers in the network.
In AP3~\cite{mislove2004ap3}, a message is forwarded over a random walk until one peer decides to send the request to the destination.
Each hop is chosen based on the lookup of a random key in a DHT, which can be misused~\cite{mittal2012information}.
In Rumor Riding~\cite{liu2011rumor} and Garlic Cast~\cite{qian2016garlic}, two random walks are started.
The random walk ends when both paths meet.
Clover~\cite{franzoni2022clover} uses a random walk to broadcast a message to the network.
The next hop is chosen based on the connection type, a message is propagated only to peers with the same connection type.
The authors distinct connections based on initiator of the connection: inbound or outbound connections.
Dandelion~\cite{bojja2017dandelion} and Dandelion++\cite{fanti2018dandelion++} offer formal anonymity guarantees for message broadcasts.
The random walk uses a privacy-subgraph for forwarding messages.
The privacy of Dandelion and Dandelion++ was investigated by~\Citeauthor{sharma2022anonymity}~\cite{sharma2022anonymity}.
Clover, Dandelion and Dandelion++ do not need an answer to the message.

RaWa-Bitswap also uses a random walk to select the proxy.
The random walk is based on the privacy-subgraph used in Dandelion/Dandelion++.
Our method focuses on content discovery in Bitswap and requires the proxy to be able to route responses back, along the requests' random walks.

In the context of research of IPFS, there exists various research investigating IPFS's
network~\cite{henningsen2020mapping,daniel2022passively},
performance~\cite{shen2019understanding, trautwein2022design},
and its subcomponents~\cite{monteiro2022enriching, kanemitsu2023lookup, sridhar2023content}.
A part of the research of IPFS, specifically focus on Bitswap.
\Citeauthor{balduf2021monitoring}~\cite{balduf2021monitoring} showed the privacy problems of Bitswap and its easy exploitation for monitoring peers interests.
While \Citeauthor{de2021accelerating}~\cite{de2021accelerating} mainly focus on performance improvement for Bitswap, their proposed method, forwarding of requests, also provides a privacy improvement.
Forwarding already provides some plausible deniability, albeit easily circumvented through passive observations.
An improvement to pure forwarding was proposed in~\cite{daniel2023improving}, where the authors introduce trickling, a diffusion spreading, with the aim to obfuscate the source.
Forwarding and trickling, both introduce a considerable load on the network although lower in the latter.
An alternative approach to improve the privacy in Bitswap was proposed in~\cite{daniel2024exploring}.
Instead of concealing the source of a request, the authors focus on concealing the content of a request.
The concealment method can introduce costly computation due to the usage of cryptographic methods.

RaWa-Bitswap makes a clear distinction between discovery and retrieval.
The discovery uses a random walk, while the retrieval is performed through a direct connection.
Gnutella 0.4~\cite{taylor2005gnutella} makes a similar distinction where \texttt{Query} messages are broadcast through the network and a \texttt{QueryHit} is routed back with information about content providers.
In general, the broadcast of messages as used in Gnutella, forwarding~\cite{de2021accelerating},
or the trickling approach proposed in~\cite{daniel2023improving}
impose a high load on the network.
RaWa-Bitswap improves the privacy, while having a much smaller footprint on the overall network load.
Although, since the dissemination of the request involves only a subset of peers compared to a broadcast, the retrieval time might be lower.
Compared to the query obfuscation proposed in~\cite{daniel2024exploring}, RaWa-Bitswap provides plausible deniability in interests themselves.
Query obfuscation still exposes the interest in something, while in RaWa-Bitswap a request might be forwarded on behalf of other nodes.
From a performance perspective the method is computation heavier and scales poorly with the number of stored blocks by a peer.
RaWa-Bitswap's performance penalty only stems from network latency and the probabilistic length of the path.

\section{Content Discovery with Random Walks}\label{sec:method}
IPFS allows the discovery and exchange of data over a P2P overlay network.
The process is handled by the Bitswap protocol.
In Bitswap, content is split into blocks, which are identified via an immutable identifier, the content identifier~(CID).
The CID consists of codec information and a multi-hash, which contains a hash, the digest length, and the used hash function.
Multiple blocks belonging to a single structure, \ie file or directory, are linked through a Merkle Directed Acyclic Graph~(DAG).
The Merkle DAG is constructed bottom-up, starting from the leaves up until a root block.
The CID of the root block is the root CID.
To be able to retrieve a file, the peer needs to know the root CID.

Content is discovered by querying all neighbors for the CID or by using a fallback system a DHT.
The discovery process reveals the interest in a CID to many unconcerned peers.
Although, the CID is only a pseudonymous representation of content, it can be used to retrieve content, which reveals the interest in the specific data,
and hence poses a privacy risk~\cite{balduf2021monitoring}.

In the following, we first provide an overview of our general approach.
Afterwards, we explain the functionality of the default Bitswap protocol before providing a more detailed explanation of RaWa-Bitswap, its subcomponents, parameters, and a discussion of some design decisions.
For the terminology, we denote a peer interested in the content as requester,
an intermediate peer relaying request due to the random walk as relay,
a peer which conducts the lookup on behalf of a peer as proxy,
and a peer in possession of the content as content provider.

\subsection{Overview}
In order to improve the privacy of Bitswap,
it is necessary to conceal the interest or the origin of a request.
Our privacy-enhancement aims to improve the privacy of the discovery process, by using a random peer of the network as a proxy.

Forwarding the message through the random walk increases the difficulty for attackers to trace the origin of a message.
By introducing message forwarding with random walks, each peer sending a message gains plausible deniability, since a message can be sent due to the random walk.
One prime example for the utilization of random walks to increase the privacy in P2P networks is Dandelion~\cite{bojja2017dandelion}.
The random walk of Dandelion and its follow-up Dandelion++~\cite{fanti2018dandelion++} are the base for our method, a Random-Walk-Bitswap~(RaWa-Bitswap).
In contrast to Dandelion/Dandelion++, which uses the random walk to broadcast a message, Bitswap also requires a response from the proxy to complete content retrieval.

\begin{figure}
    \resizebox{\columnwidth}{!}{
    \begin{tikzpicture}[>=stealth]
        \tikzstyle{server}=[circle, draw, thin,fill=sron0!40, minimum width=8mm]
        \tikzstyle{client}=[circle, draw, thin,fill=sron1!40, minimum width=8mm]
        \tikzstyle{link}=[thick, dotted]

        \node [client] (a) at (0.25, 6.75) {C};
        \node [client] (b) at (1.25, 8.75) {R};
        \node [client] (c) at (1.50, 4.25) {};
        \node [client] (d) at (2.00, 5.50) {};
        \node [client] (e) at (3.25, 9.50) {R};
        \node [client] (f) at (3.00, 7.50) {R};
        \node [client] (g) at (4.00, 5.50) {CP};
        \node [client] (h) at (3.75, 3.25) {};
        \node [client] (i) at (5.50, 7.25) {P};
        \node [client] (j) at (6.00, 10.0) {};
        \node [client] (k) at (6.50, 5.50) {};
        \node [client] (l) at (6.50, 3.00) {};
        \node [client] (m) at (8.00, 8.50) {};
        \node [client] (n) at (9.50, 3.75) {};
        \node [client] (o) at (8.50, 6.00) {};
        \node [client] (p) at (10.0, 6.75) {};
        \node [client] (q) at (9.75, 9.75) {};

        \node[draw=none] at (5.50, 2.00) {Phases};
        \draw[<-, ultra thick, sron0] (1.00, 1.25) -- node[label,above, midway]{Privacy} (2.25, 1.25);
        \draw[<-, ultra thick, sron1] (3.50, 1.25) -- node[label,above, midway]{Proxy} (4.75, 1.25);
        \draw[<-, ultra thick, sron2] (6.00, 1.25) -- node[label,above, midway]{Return} (7.25, 1.25);
        \draw[<->, ultra thick, sron3] (8.50, 1.25) -- node[label,above, midway]{Exchange} (9.75, 1.25);

        \node[draw=none, sron0, font=\bfseries] at (2.00, 8.00) {1. Privacy};
        \path[->, ultra thick, sron0, transform canvas={yshift=-0.5mm, xshift=1mm}] (a) edge (b);
        \path[->, ultra thick, sron0, transform canvas={yshift=-1mm, xshift=0.5mm}] (b) edge (e);
        \path[->, ultra thick, sron0, transform canvas={yshift=0.25mm, xshift=-1mm}] (e) edge (f);
        \path[->, ultra thick, sron0, transform canvas={yshift=-1mm, xshift=-0.00mm}] (f) edge (i);

        \node[draw=none, sron1, font=\bfseries] at (6.66, 8.50) {2. Proxy};
        \path[->, ultra thick, sron1, transform canvas={yshift=1mm, xshift=1mm}] (i) edge (e);
        \path[->, ultra thick, sron1, transform canvas={yshift=0.0mm, xshift=1mm}] (i) edge (j);
        \path[->, ultra thick, sron1, transform canvas={yshift=0.5mm, xshift=1mm}] (i) edge (k);
        \path[->, ultra thick, sron1, transform canvas={yshift=1mm, xshift=-0.25mm}] (i) edge (m);
        \path[->, ultra thick, sron1, transform canvas={yshift=1mm, xshift=0.5mm}] (i) edge (o);
        \path[<->, ultra thick, sron1, transform canvas={yshift=-1mm, xshift=1mm}] (i) edge (g);

        \node[draw=none, sron2, font=\bfseries] at (1.50, 9.50) {3. Return};
        \path[<-, ultra thick, sron2, transform canvas={yshift=0.5mm, xshift=-1mm}] (a) edge (b);
        \path[<-, ultra thick, sron2, transform canvas={yshift=1mm, xshift=-0.5mm}] (b) edge (e);
        \path[<-, ultra thick, sron2, transform canvas={yshift=-0.25mm, xshift=1mm}] (e) edge (f);
        \path[<-, ultra thick, sron2, transform canvas={yshift=1mm, xshift=-0.00mm}] (f) edge (i);

        \path[<->, ultra thick, sron3] (a) edge node [midway, above, sloped, font=\bfseries] {4. Exchange} (g);

        \path[link] (a) edge (b) edge (c) edge (f) edge (d);
        \path[link] (b) edge (e) edge (f) edge (d);
        \path[link] (c) edge (d) edge (g) edge (h);
        \path[link] (d) edge (f) edge (g) edge (h);
        \path[link] (e) edge (f) edge (i) edge (j) edge (m);
        \path[link] (f) edge (g) edge (i) edge (j) edge (k);
        \path[link] (g) edge (h) edge (k) edge (l) edge (i);
        \path[link] (h) edge (k) edge (l);
        \path[link] (i) edge (j) edge (k) edge (m) edge (o); 
        \path[link] (j) edge (m) edge (q);
        \path[link] (k) edge (m) edge (o) edge (l) edge (n);
        \path[link] (l) edge (n) edge (o);
        \path[link] (m) edge (o) edge (p) edge (q);
        \path[link] (n) edge (o) edge (p);
        \path[link] (o) edge (p);
        \path[link] (p) edge (q);
    \end{tikzpicture}
    }
    \caption{Phases of RaWa-Bitswap.}
    \label{fig:network}
\end{figure}
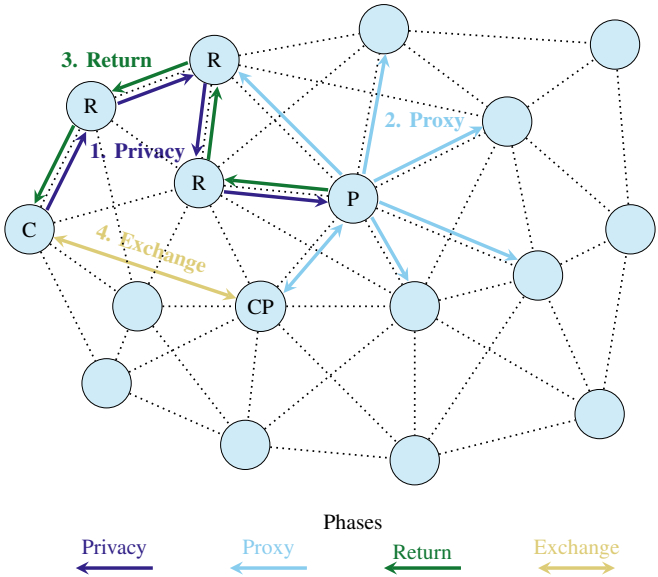

The main goal of RaWa-Bitswap is to enhance privacy of content requests in general.
Therefore, the protection method mainly aims to protect against passive listeners and not active traffic manipulating adversaries, or targeted attacks.
RaWa-Bitswap is split into four phases: privacy-phase, proxy-phase, return-phase, exchange-phase.
\cref{fig:network} provides a rough overview of our method.
In the privacy phase the proxy is selected by utilizing a random walk.
In a random walk, a neighboring peer is selected at random to which the message or request is relayed.
The selected peer decides based on a probability to either become a proxy or to continue the walk.
If the walk continues, the selection process is repeated and the relay randomly selects a new peer.
Through this process the random walk eventually terminates at a random peer, which becomes the proxy.
At the proxy the protocol enters the proxy-phase.
In the proxy phase the proxy executes the normal Bitswap discovery by querying all neighbors.
Once the proxy finds a content provider, RaWa-Bitswap enters the return-phase.
In the return phase, the proxy returns information about the content provider(s) via the path of the random walk to the requestor.
After the requester received a content provider, the exchange phase begins.
In the exchange phase, the requester connects to a content provider and requests the data.

\subsection{Vanilla-Bitswap}
The default Bitswap protocol, which we denote here as Vanilla-Bitswap, uses different request and response types, which are wrapped in the same envelope, a Bitswap message.
The requests are \texttt{WANT-HAVE}, \texttt{WANT-BLOCK}, and \texttt{CANCEL}, and the responses are \texttt{HAVE}, \texttt{DONT-HAVE}, and \texttt{BLOCK}.
Content retrieval follows a pattern using the requests.
First, it is necessary to identify content providers.
The content discovery is handled by the \texttt{WANT-HAVE} request.
A peer sends to all directly connected peers, its neighbors, a \texttt{WANT-HAVE} message, which is answered with a \texttt{HAVE} in case the neighbor stores the block, or can be answered with a \texttt{DONT-HAVE}, in case the neighbor does not store the block.
If a content provider is found, the block is requested with a \texttt{WANT-BLOCK}, which the content provider answers with \texttt{BLOCK}.
Once the block is retrieved, all peers, which received a \texttt{WANT-HAVE} of the block, receive a \texttt{CANCEL}.
In summary, content is discovered by announcing the interest with a \texttt{WANT-HAVE} to all neighbors and satisfaction or disinterest in content is announced with \texttt{CANCEL}.
In case none of the neighbors can provide the content, Bitswap needs a fallback subsystem to find new peers which might store the content.
IPFS uses a Kademlia-based DHT as the fallback subsystem to search for content providers, which is queried after a timer~$t_{1}$ expires.

\subsection{RaWa-Bitswap}
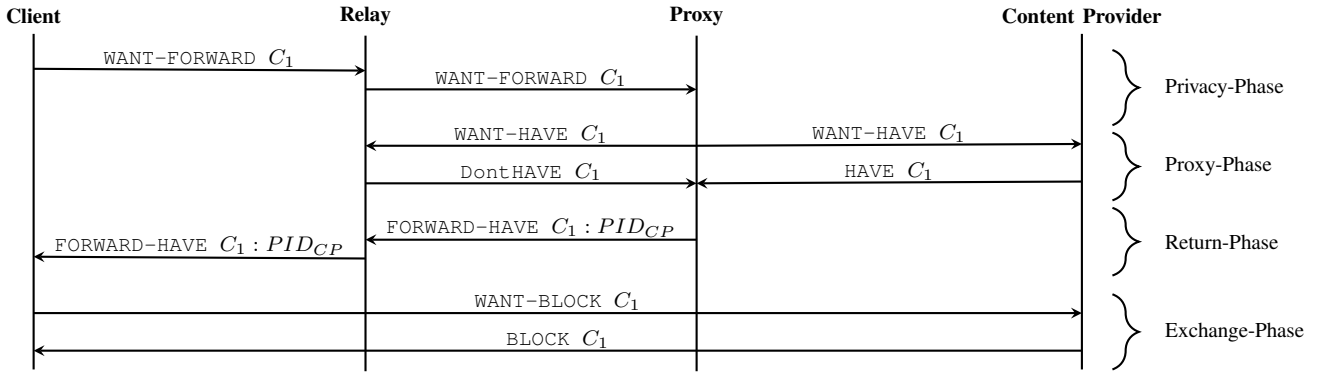
\begin{figure*}\footnotesize
      \begin{tikzpicture}[node distance=35mm,auto,>=stealth]
        \tikzstyle{label}=[above,yshift=-2pt,font=\tt\footnotesize,scale=1]

        \node[font=\bf\footnotesize] (a) {Client};
        \node[font=\bf\footnotesize,right= of a] (b) {Relay};
        \node[font=\bf\footnotesize,right= of b] (c) {Proxy};
        \node[font=\bf\footnotesize,right= of c] (d) {Content Provider};
        \coordinate[below=50mm of a.south] (a_ground);
        \coordinate[below=50mm of b.south] (b_ground);
        \coordinate[below=50mm of c.south] (c_ground);
        \coordinate[below=50mm of d.south] (d_ground);

        \draw[thick] ([yshift=-1pt]a.south) -- ([yshift=5mm]a_ground);
        \draw[thick] ([yshift=-1pt]b.south) -- ([yshift=5mm]b_ground);
        \draw[thick] ([yshift=-1pt]c.south) -- ([yshift=5mm]c_ground);
        \draw[thick] ([yshift=-1pt]d.south) -- ([yshift=5mm]d_ground);
        \draw[->,thick] ($(a.south)!0.10!(a_ground)$) -- node[label,midway]{WANT-FORWARD $C_{1}$} ($(b.south)!0.10!(b_ground)$);
        \draw[->,thick] ($(b.south)!0.15!(b_ground)$) -- node[label,midway]{WANT-FORWARD $C_{1}$} ($(c.south)!0.15!(c_ground)$);

        \draw[->,thick] ($(c.south)!0.30!(c_ground)$) -- node[label,midway]{WANT-HAVE $C_{1}$} ($(d.south)!0.30!(d_ground)$);
        \draw[->,thick] ($(c.south)!0.30!(c_ground)$) -- node[label,midway]{WANT-HAVE $C_{1}$} ($(b.south)!0.30!(b_ground)$);
        \draw[<-,thick] ($(c.south)!0.40!(c_ground)$) -- node[label,midway]{HAVE $C_{1}$} ($(d.south)!0.40!(d_ground)$);
        \draw[->,thick] ($(b.south)!0.40!(b_ground)$) -- node[label,midway]{DontHAVE $C_{1}$} ($(c.south)!0.40!(c_ground)$);

        \draw[<-,thick] ($(b.south)!0.55!(b_ground)$) -- node[label,midway]{FORWARD-HAVE $C_{1}:PID_{CP}$} ($(c.south)!0.55!(c_ground)$);
        \draw[<-,thick] ($(a.south)!0.60!(a_ground)$) -- node[label,midway]{FORWARD-HAVE $C_{1}:PID_{CP}$} ($(b.south)!0.60!(b_ground)$);

        \draw[->,thick] ($(a.south)!0.75!(a_ground)$) -- node[label,midway]{WANT-BLOCK $C_{1}$} ($(d.south)!0.75!(d_ground)$);
        \draw[<-,thick] ($(a.south)!0.85!(a_ground)$) -- node[label,midway]{BLOCK $C_{1}$} ($(d.south)!0.85!(d_ground)$);

        \draw [decorate,decoration={brace,amplitude=10pt,raise=40pt},thick]
        ([xshift=-10mm]$(d.south)!0.05!(d_ground)$) -- ([xshift=-10mm]$(d.south)!0.25!(d_ground)$)
        node [midway,text width=60mm,xshift=20mm, font=\footnotesize] {Privacy-Phase};

        \draw [decorate,decoration={brace,amplitude=10pt,raise=40pt},thick]
        ([xshift=-10mm]$(d.south)!0.27!(d_ground)$) -- ([xshift=-10mm]$(d.south)!0.45!(d_ground)$)
        node [midway,text width=40mm,xshift=20mm, font=\footnotesize] {Proxy-Phase};

        \draw [decorate,decoration={brace,amplitude=10pt,raise=40pt},thick]
        ([xshift=-10mm]$(d.south)!0.47!(d_ground)$) -- ([xshift=-10mm]$(d.south)!0.65!(d_ground)$)
        node [midway,text width=40mm,xshift=20mm, font=\footnotesize] {Return-Phase};

        \draw [decorate,decoration={brace,amplitude=10pt,raise=40pt},thick]
        ([xshift=-10mm]$(d.south)!0.70!(d_ground)$) -- ([xshift=-10mm]$(d.south)!0.90!(d_ground)$)
        node [midway,text width=40mm,xshift=20mm, font=\footnotesize] {Exchange-Phase};

      \end{tikzpicture}
    \caption{Message sequence of RaWa-Bitswap.}
    \label{fig:rawa-swap}
 \end{figure*}

In RaWa-Bitswap, a random walk is prepended to the discovery process of Vanilla-Swap, resulting in a proxy executing the discovery of content providers.
The proxy executes only content discovery (\texttt{WANT-HAVE}) and returns the discovered content providers back over the random path of the random walk.
The requester executes the content retrieval (\texttt{WANT-BLOCK}) by itself by connecting to the provided content providers.
The phases and message sequence with only one relay node is shown exemplary in \cref{fig:rawa-swap}.
Since each peer probabilistically decides whether to relay or become a proxy, it is possible that the first peer after the requester already becomes the proxy.
Respectively, it is also possible that there are more peers between requester and proxy.

In the privacy-phase, instead of sending a \texttt{WANT-HAVE} to all neighbors, the peer starts a random walk.
For the random walk, we introduce a new request type a \texttt{WANT-FORWARD}, indicating the relay of the lookup.
Therefore, the requester sends a \texttt{WANT-FORWARD} request for the interested CID to a selected neighbor.
Each peer relays the \texttt{WANT-FORWARD} request with probability~$1-p$ or transitions into the proxy phase with probability~$p$.
The relaying peers store the CID, the predecessor, and successor of a request.
The proxy only stores the predecessor and CID of the request.
In the proxy phase, the peer executes the Vanilla-Bitswap lookup, sending a \texttt{WANT-HAVE} to all neighbors and after a time-out doing a DHT lookup for the item.
After the proxy discovered content provider(s), the return phase begins.

In the return phase, a list with the content providers is sent back with a \texttt{FORWARD-HAVE} response along the path of the random walk to the original requester.
The list contains only the peer identifier of the content provider.
In case the proxy is aware of the contact address of a content provider, the information is also included in the list.
Once the original peer receives the list of content providers, the exchange phase begins.
During the exchange phase, the peer selects one content provider to connect to and requests the block with a \texttt{WANT-BLOCK}.
As a fallback solution for the content discovery, the requester has a fallback search timer~$u$.
After the timer times out, the requester performs a DHT lookup.

Due to the dependence on other peers, there is an additional functionality to make the protocol more resilient to churn, \ie joining and leaving of peers.
Churn is a problem, since a relay may depart from the network before the discovery process is finished.
If a relay departs the response of a proxy can no longer reach the requester.
To reduce the impact of incomplete walks, the requester periodically re-transmits the request based on a timer~$t_{0}$.
A relay receiving the same CID of a predecessor, directly re-transmits the request to the same successor.
If the successor left the network and is no longer reachable, the relay directly transitions into the proxy phase, instead of selecting a new peer from the privacy-subgraph.
Therefore, a re-transmit results in the same path, although the path might be shorter, due to absent successor.
This ensures that the path is complete and missing responses are a result of lacking content providers.

\subsection{Privacy-Subgraph}\label{sub-sec:subgraph}
The random walk serves two purposes, it selects a random peer from the network as a proxy, and provides plausible deniability for each peer during the selection process.
The selection of the next peer needs some considerations.

In RaWa-Bitswap, the next peer is selected based on a privacy-subgraph.
The usage of a privacy-subgraph for selecting the next peer was introduced in Dandelion and Dandelion++ to increase request mixing.
The aim is that requests from different requesters move along the same link(s), increasing the difficulty for adversaries to find the source of a request.
The privacy-subgraph is a directed graph with a specific in- and out degree.
In case of Dandelion the privacy-subgraph is a 2-reqular graph (1~successor, 1~predecessor).
For Dandelion++, the privacy-subgraph was changed to a directed 4-regular graph (2 predecessors, 2 successors), allowing the peer to choose between two peers for the message relay.

The construction of the privacy-subgraph in RaWa-Bitswap is based on the construction method of Dandelion++, with the difference of using a configurable out-dgree $\eta$.
The construction is asynchronous and non-interactive.
For each neighbor, a set of $\eta$ neighbors are uniformly at random selected as successors.
Therefore, a graph is generated in which each node has a set of $\eta$ directed edges.
The constructed graph is not an exact regular graph, but its expected node degree is $d = 2\eta$.

Dandelion's privacy guarantees depend on a private privacy-subgraph, relaxed in Dandelion++ with the higher node degree, and investigated in \cite{sharma2022anonymity}.
The authors of \cite{sharma2022anonymity} propose a Bayesian framework for evaluating the anonymity of P2P network schemes and applied it to Dandelion and Dandelion++.
By using the entropy of potential request origins as an anonymity metric, the authors found that the anonymity of both protocols is limited.
They discovered that increasing the node degree leads to better anonymity.
Furthermore, the authors assume the peer's privacy-subgraph is known.
In general, the privacy-subgraph can be learned by connecting to all peers and sending many requests to one peer.
Based on the propagation of requests, the privacy-subgraph can be estimated.
However, the frequent reconstruction of the privacy-subgraph in Dandelion/Dandelion++ is not considered in \cite{sharma2022anonymity}.

As a consequence, we periodically reconstruct the privacy-subgraph and make the out-degree configurable.
The concrete value of the reconstruction timer $r$ depends on the expected load at the peer and should be adjusted accordingly.
As an initial value, we set $r$ to $540\,s$ as used in Dandelion.

\begin{table}
  \caption{Parameters of RaWa-Bitswap.}
  \label{tab:para}
  \footnotesize
  \begin{tabularx}{\columnwidth}{llr}
    \toprule
     Parameter & Description & Value \\
     \midrule \midrule
     $p$ & proxy transition & configurable \\
     $t_{0}$ & re-transmit \texttt{WANT-FORWARD} (requester) & $1\,s$ \\
     $t_{1}$ & timer for fallback lookup (proxy) & $1\,s$ \\
     $u$ & timer for fallback lookup (requester) & configurable \\
     $\eta$ & privacy-subgraph out degree & configurable \\
     $r$ & privacy-subgraph reconstruction timer & $540\,s$ \\
    \bottomrule
  \end{tabularx}
\end{table}

\subsection{Parameter}
RaWa-Bitswap has some properties that can be adjusted with parameters.
\Cref{tab:para} provides an overview of all parameters used in RaWa-Bitswap.
The value of these parameter needs to be considered for the privacy-utility trade-off of RaWa-Bitswap.
In the following, we focus on $p$ and the timers $t_{0}$, $t_{1}$, and $u$.
The parameters $\eta$ and $r$, relevant for the privacy-subgraph, have been discussed in the previous section.

The value of the proxy transition probability $p$ decides the approximate path length of the walk.
In general, a lower value of $p$ decreases performance, while increasing the privacy, due to the longer path.
The probability $X_{p}$ that a peer becomes a proxy within $e$ hops with $p$ can be calculated by:
\begin{align}
X_{p}(e) = 1-(1-p)^e.
\end{align}
This means with $p=0.2$, $90\,\%$ of the paths have a length $\le 11$ and with $p=0.3$, $90\,\%$ of the paths have a length $\le 9$.
The authors of Dandelion++ evaluated that values for $p \le 0.2$ have only a limited impact on the privacy.
Although higher values have a stronger impact on the privacy exposure, the improvement of the performance might justify higher values.

The re-transmit timer~$t_{0}$ mitigates the impact of churn on the lookup process.
Lower values increase resilience against churn and increase the load on the network.
The value depends on the network behavior.
We propose to use $t_{0}=1\,s$.

The proxy fallback timer~$t_{1}$ is the same timer as used in Vanilla-Bitswap for querying the fallback system, the DHT.
In Vanilla-Bitswap, this timer is set to $1\,s$, at the time of writing.

The requester fallback lookup timer~$u$ is an additional backup solution in case the proxy lookup takes too much time.
The fallback lookup by the requester should be avoided to increase the privacy, but can speed up lookup times in case of long delays.
The value of $u$ should be at least higher than the fallback timer of the proxy $t_{1}$ and the re-transmit timer~$t_{2}$.
We propose to set $u=2 \cdot t_{0}$, which is $2\,s$.

\subsection{Discussion}\label{sub-sec:vulnerability}
In RaWa-Bitswap, the proxy only executes content discovery.
The content providers are routed back to the requester.
As an alternative, the proxy could also retrieve blocks, relaying the \texttt{BLOCK} responses.
A \texttt{BLOCK} response is compared to a \texttt{HAVE} very large with kilobytes compared too few hundreds of bytes.
This increases the load on the network and further increases the time of message transmission.
The newly introduced \texttt{FORWARD-HAVE} could also be large for popular files with many content providers, however, in general, the size is also low.
Excluding content retrieval reduces the load on the network and reduces RaWa-Bitswap's susceptibility to churn.
The lower load reduces the round trip time of content discovery, which reduces the risk that an intermediate peer leaves the network before the content discovery is finished.

Another aspect is that files can be divided into multiple blocks, which are likely to be located at the same peer.
Once a content provider for the root block is discovered, the rest of the blocks may also be retrieved from the same peer.
If the proxy only executes content discovery, the requester learns the content provider and can request followup blocks directly.

Focusing only on content discovery also presents its own challenges.
The interest in a CID is still revealed through \texttt{WANT-BLOCK} requests.
In RaWa-Bitswap, a \texttt{WANT-BLOCK} request is exclusively sent by a requestor.
Therefore, plausible deniability is lost as soon as the requester sends the \texttt{WANT-BLOCK} request.
This is deemed acceptable, as the protocol's objective is to ensure network-wide privacy, rather than focusing on the individual level.
However, this behavior can be exploited to find the origin of a request.
A peer receiving a \texttt{WANT-FORWARD} immediately responds with a \texttt{FORWARD-HAVE} that contains only itself or a peer controlled by the attacker.
As a result, the attacker will receive a \texttt{WANT-BLOCK} request for the item from the original requester.
In this case the attacker does not even need to store the block.
The identification risk can be reduced, if the requester keeps the Vanilla-Bitswap behavior of first sending a \texttt{WANT-HAVE} request, simulating a proxy.
This additional verification does not need to be sent to all neighbors but only to the newly connected content provider.
In this case the attacker could only learn the requester, if it also stores the block.
However, nothing prevents the attacker from also sending false \texttt{HAVE} responses.
The method can be enhanced with alternative methods, which obfuscate the item of a request \cite{daniel2024exploring}, further reducing the risk.

\section{Implementation Details and Integration}\label{sec:impl}
RaWa-Bitswap requires some modifications to Bitswap
as well as some new additions, \ie,
a requester session, a relay manager, a proxy session, and a forward graph manager.
All newly introduced parameters
are configurable when instantiating RaWa-Bitswap.
A PoC implementation based on Bitswap of the \texttt{boxo} library~\cite{gitgoboxo} v0.8.0 is publicly available.%
\footnote{https://github.com/manuelwedler/boxo}

\subsubsection*{Bitswap Modifications}
In Bitswap, blocks below $1024\,B$ are immediately sent in response to \texttt{WANT-HAVE} requests.
This is removed from the RaWa-Bitswap implementation, since only a proxy sends \texttt{WANT-HAVE} requests and might not need the block.
Furthermore, we added a new request and its respective response to the Vanilla-Bitswap messages: \texttt{WANT-FORWARD} and \texttt{FORWARD-HAVE}.
The \texttt{WANT-FORWARD} indicates the relay of the lookup.
The aim of the \texttt{WANT-FORWARD} is to be able to distinguish between answering requests and forwarding requests.
Although, this can also be accomplished by using a flag into \texttt{WANT-HAVE} requests, the new request type makes the protocol more comprehensible.
The \texttt{FORWARD-HAVE} is the response to a \texttt{WANT-FORWARD}, transmitting the outcome of the content discovery.
Since a \texttt{HAVE} only contains the CID and implies that the sender stores a block, a new response type is necessary.
The new request and response utilize the same Bitswap message envelope.

\subsubsection*{Requestor Session}
The requester session is primarily the Vanilla-Bitswap session.
The difference is that whenever a \texttt{WANT-HAVE} request would be sent to peers, instead a \texttt{WANT-FORWARD} request for the CID is sent to one peer.
Instead of waiting for \texttt{HAVE} responses, the session expects \texttt{FORWARD-HAVE} responses.
The session also re-transmits requests in case no response is received after a time-out.

\subsubsection*{Relay Manager}
The relay manager deals with the random relay of \texttt{WANT-FORWARD} and the correct return of the \texttt{FORWARD-HAVE} response.
The relay manager is based on the implementation of~\cite{de2021accelerating}.
Upon receiving a new \texttt{WANT-FORWARD} request, the peer stores predecessor and CID of the request.
Afterwards, it starts a proxy session with probability~$p$ or becomes a relay with probability~$1-p$.
As a relay, the peer stores the successor and relays the request.
A relay, receiving a request from the same source for the same CID, forwards the request to the same successor.
If a successor disconnected from the relay, the relay transitions into the proxy phase.
In case of a request from a different source with the same CID, a peer is selected, which has not yet received a request for the CID from this relay, reducing relay loops.

\subsubsection*{Proxy Session}
A proxy session searches and returns content providers to the relay manager.
The proxy session primarily behaves as a Vanilla-Bitswap session.
However, a proxy session only handles single CIDs, checks the local storage for the block, and does not send \texttt{WANT-BLOCK} requests.

\subsubsection*{Forward Graph Manager}
The \texttt{forwardGraphManager} structure is introduced into the \texttt{peermanager} package.
The \texttt{forwardGraphManager} selects succeeding peers for the privacy-subgraph.
The number of peers depends on a specified target out-degree.
The \texttt{forwardGraphManager} selects peers from the privacy-subgraph according to a pre-configured strategy, defined in the \texttt{forwardstrategy} package.

\section{Evaluation}\label{sec:eval}
We evaluate RaWa-Bitswap's privacy gains
and its impact on performance with a code simulation
using Testground v0.6.0.%
\footnote{https://github.com/testground/testground (Accessed: 2024-02)}
Testground is a testing platform
that enables code execution with a simulated network.
The scalability is constrained by the available hardware.
For our simulations, we used an Intel Core i7-6700K CPU with $16\,GiB$ RAM and a 64-bit Ubuntu 22.04.3 LTS system with Linux kernel 6.2.0-33-generic as the operating system.
The code simulations are defined in a testplan, which is publicly available.%
\footnote{https://github.com/manuelwedler/boxo/tree/main/testplans/rawa-bitswap}

\subsection{Simulation Environment}
All simulations use the same principle to build a network consisting of $50$~peers.
Due to hardware constrains, networks with more peers were not simulated.
Each peer establishes a connection to four randomly selected other peers.
The selection chooses only peers, which are not connected to the peer.
Therefore, a node has at least four outgoing connections and the node degree is on average eight.
The links are configured with a latency of $100\,ms$, a $10\,ms$ jitter, and a bandwidth of $1\,MiB/s$.
Bitswap runs on top of a \texttt{libp2p}\footnote{https://github.com/libp2p/go-libp2p (Accessed: 2024-02)} host with a TCP and QUIC address.
To simplify the simulation, we use a dummy DHT as the content routing subsystem.
The dummy DHT knows all block locations, resolving all queries after a delay of $622\,ms \pm 10\,\%$.
This delay is based on the median duration of a DHT query in IPFS as determined in~\cite{trautwein2022design}.

All simulations execute the same behavior.
Each honest node generates and stores one unique, random block with a predetermined size.
Afterwards, the honest nodes concurrently query Bitswap for the CID of a block from another node.
A single CID might be queried multiple times, introducing some replication and the possibility for multiple content providers.
In IPFS, blocks are initially available from a single source
and replicated only upon request.
Therefore, our simulations reflect the initial state of blocks.

We run the simulations using different parameters.
Each parameter combination is run 100 times.
The parameters $t_{0}=1\,s$, $t_{1}=1\,s$, $u=2\,s$, and $r=540\,s$ are fixed for all simulations.
Due to the high value of $r$ and the execution of only one retrieval round, the privacy-subgraph stays the same during one simulation run.
For the privacy-subgraph, we evaluate the values of $\eta=1$ (Dandelion), $\eta=2$ (Dandelion++), and $\eta=max$, which means the privacy-subgraph consists of all neighbors.

\input{priv-plot-100-1-all.tex}

\subsection{Privacy Evaluation}
We quantify the degree of network-wide privacy that RaWa-Bitswap achieves by introducing a classification problem.
An adversary seeks to link any observed CID to a peer that is interested in the block.
Therefore, the classification problem is the peers interested in CIDs.
We quantify the success using precision and recall.
Based on the acquired information, the adversary tries to link a CID to all honest peers.
To determine network wide privacy, the individual precision and recall are averaged.
The adversary knows all peers of the network, controlling a fraction of peers, denoted with $\alpha$.
While our method aims to improve the privacy against passive adversaries, we also show the privacy-enhancement against active adversaries.
Therefore, we assume three adversaries with different capabilities: First-spy estimator~(FSE), \texttt{WANT-FORWARD} Exploiter~(WFE), and subgraph-aware WFE~(SAWFE).

The FSE is the simplest adversary controlling only one node, $\alpha=0.02$.
However, the node of the FSE connects to all other peers.
The FSE is a passive adversary, assuming the interest of a peer is the CID of the first request seen from the peer.
In case a peer never became a proxy or never forwarded requests to the malicious node, a peer cannot be linked to a CID.
The peers to which a CID could not be assigned, are linked to an observed CID at random.
Observed CIDs are in this context any CID for which the FSE saw a request.

The WFE exploits the vulnerability mentioned in \Cref{sub-sec:vulnerability}.
This adversary tries to actively exploit RaWa-Bitswap to reveal the origin of requests.
The aim of the WFE is to trigger a requester to send a \texttt{WANT-BLOCK} request to any controlled node by claiming to be a content provider.
The WFE controls 10 nodes ($\alpha=0.2$) and each node establishes four connections to distinct honest nodes.
By doing this, each honest node has at least one connection with a malicious node and the average node degree of the network remains eight.
The WFE combines the data from all its controlled nodes.
Every peer is associated with the CID of the first \texttt{WANT-BLOCK} received from that peer.
Similar to the FSE, in case no \texttt{WANT-BLOCK} request is received, the peer gets assigned an observed CID at random.
While the WFE produces a fake \texttt{FORWARD-HAVE}, it otherwise behaves like an honest node, forwarding requests and might even act as a proxy.

While the FSE and WFE have no knowledge of the privacy-subgraph, the privacy-subgraph is provided to the SAWFE.
In practice, this knowledge could be obtained by sending multiple requests as described in \Cref{sub-sec:subgraph}.
Otherwise, the SAWFE has the same capabilities as the WFE,
\ie, active, and $\alpha=0.2$.
The SAWFE also maps every peer to the CID of the first \texttt{WANT-BLOCK} observed from this peer.
The difference between SAWFE and WFE is the handling of peers for which no \texttt{WANT-BLOCK} could be observed.
All observed \texttt{WANT-HAVE} requests are observed and each CID is assigned to an unclassified direct predecessor.
Due to RaWa-Bitswap's functionality, a predecessor of a proxy must be the requester.
The resilience against predecessor attacks~\cite{wright2004predecessor}
depends accordingly on the privacy of the privacy-subgraph.

\cref{fig:priv-100-1} shows the precision and recall for all three adversaries.
In Vanilla-Bitswap, a FSE can with almost $100\,\%$ certainty determine the interest of all observed peers.
RaWa-Bitswap reduces precision and recall of the prediction down to $40\,\%$ -- $50\,\%$.
The effect of $p$ on the precision and recall seems almost negligible.
Considering $\eta$, a value of 1 or max produce almost the same results with 1 having slightly better results.
For $\eta=2$, the values are better than Vanilla-Swap but overall the worst precision and recall of on average $\approx 60\,\%$.
Still, all values are much better compared to Vanilla-Bitswap.

For the active adversaries the privacy gains are less significant.
While the precision and recall is still lower than the values for the FSE in Vanilla-Bitswap, the precision and recall is now $60\,\%$ -- $80\,\%$.
For $\eta=1$, $p$ seems to have little effect on precision and recall, which might be due to the occurrence of loops which cut the paths short.
For bigger $\eta$, a higher value of $p$ seems beneficial to the privacy.
Comparing the results for WFE and SAWFE, the SAWFE has similar precision values, although slightly worse than WFE.
The recall of the SAWFE is also similar but slightly better than the WFE.

\begin{figure*}
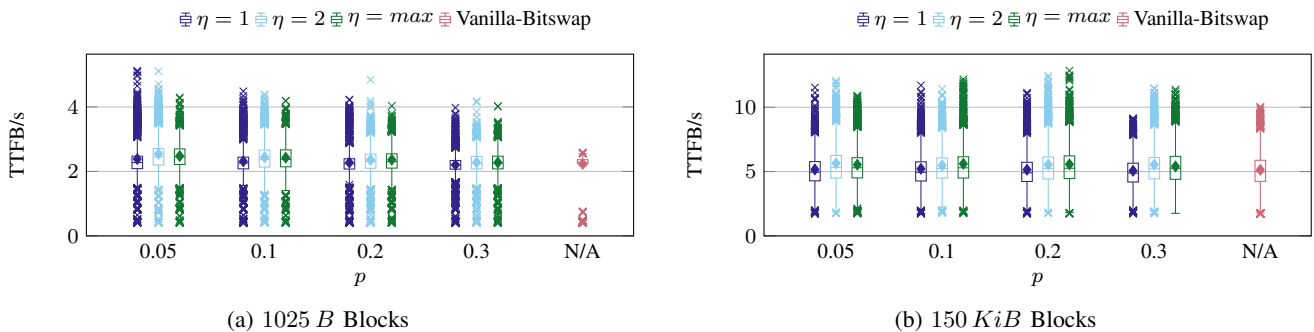

    \subcaptionbox{$1025\,B$ Blocks\label{fig:perf-small}}[0.49\textwidth]{\input{performance-plot-100-1.tex}}
    \subcaptionbox{$150\,KiB$ Blocks\label{fig:perf-large}}[0.49\textwidth]{\input{performance-plot-100-150.tex}}
    \caption{TTFB for different $p$ in comparison to Vanilla-Bitswap.}
    \label{fig:perf}
\end{figure*}

In case of WFE and SAWFE, the behavior of the privacy seems to be counterintuitive.
Longer paths should provide better privacy.
This is probably due to the nature of the exploit.
The exploit functions very well in case a malicious peer is encountered during the privacy phase, \ie, the exploit is guaranteed to succeed.
If a malicious peer is encountered during the proxy phase, there is a chance that an honest peer also sends a \texttt{HAVE} providing the proxy with multiple content providers.
Therefore, the exploit might fail.
Against WFE and SAWFE shorter paths should be beneficial, since the chance of encountering a malicious peer is lower.
The difference between the WFE and SAWFE is also due to the exploit.
The SAWFE has only an advantage in case the exploit fails and needs to guess the origin of the request.
Due to the knowledge of the predecessor of a peer, a guess has a higher chance to succeed.

In summary, RaWA-Bitswap shows clearly visible privacy improvements against passive adversaries and only a minor privacy improvement in case of the active adversary.
Our active adversaries exploit a vulnerability of the protocol.
The vulnerability can be mitigated by obfuscating the item of a request, making it harder for an adversary to use the exploit.
However, it does not fully remove the threat.
The exploit can be prevented by also using the proxy for content retrieval, although, this would increase network load.
The negligible effect of $\eta$ and $p$ might be due to the small network.

\subsection{Performance Evaluation}
We determine the performance overhead of RaWa-Bitswap with a direct comparison to Vanilla-Bitswap.
Similar to~\cite{de2021accelerating}, we use the time-to-first-block~(TTFB) as the performance metric.
TTFB is the time elapsing from (locally) sending the request to Bitswap until the first block is received.
For the simulation, every content item is encapsulated in a single block of $1025\,B$ and $150\,KiB$.
While TTFB accordingly yields the content retrieval time,
it is still reasonable to make statements on the content discovery.
Especially for smaller blocks, the download time is negligible.
\cref{fig:perf} shows the simulation results.

\cref{fig:perf-small} shows the result for a comparatively small block and the time mainly covers content discovery.
For Vanilla-Swap, we can see very low variations of the retrieval time with an average of around $2.2\,s$.
RaWa-Bitswap shows larger variations, although, a similar average of around $2.2\,s$ -- $2.4\,s$ between the different $\eta$ and $p$.
The large variations are due to the random nature of the walk.
The number of relays varies with each request.
While lower values for $p$ should result in longer paths and therefore longer TTFB, it seems that the variation of $p$ has almost no influence on the performance.
The influence of $\eta$ is slightly higher.
Bigger values for $\eta$ result in higher TTFB.
This behavior might be a result of our comparatively small network and consequently higher chance of loops.
The occurrence and influence of loops is further discussed in~\Cref{sub-sec:eval-discuss}.

The general behavior is similar for larger blocks as shown in \cref{fig:perf-large} where a larger block is retrieved.
The TTFB is in general higher, due to the increased amount of data that needs to be retrieved.
For larger blocks the content discovery time has a lower influence on the whole retrieval time.
In summary, the simulation shows that the impact of the random walk on the TTFB is rather small.

\subsection{Discussion}\label{sub-sec:eval-discuss}
The results concerning the behavior for different values of $p$, could be due to our limited number of nodes and the occurrence of loops.
The privacy-subgraph construction only approximates a regular graph, incorporating loops due to nodes which choose each other as successors.
In case of a loop, it is possible that a node can no longer choose a successor for the message, which results in the immediate transition into the proxy phase.
This is especially notable for $\eta=1$, where each peer only has one successor.
Loops reduce the impact of $p$, since the random walk finishes earlier.
The occurrence of loops reduces with the size of the network.

Loops can have an influence on privacy and performance.
For the privacy, shorter paths impact the gained privacy.
The FSE is less likely to be affected, since it is connected to all nodes anyway and therefore the path length has no significant influence on the discovery.
Only if the first chosen peer is the FSE, the guess is correct.
Against the WFE and SAWFE, a shorter path reduces the probability to encounter a malicious peer.
However, the SAWFE also benefits from a shorter path in case the exploit does not work.
Since the SAWFE knows the predecessors of a relay, it can more reliably guess the origin of a request.
For performance, a shorter random walk means earlier completion of content discovery, which results in a shorter TTFB.
As a result, the impact of RaWa-Bitswap on the performance of content retrieval might be larger in networks with more peers.
While the small network limits the performance evaluation, it has only a low impact on the general privacy results.
A passive observer can comparatively easy connect to many peers, receiving the results of the FSE for Vanilla-Bitswap.
Therefore, against a FSE, RaWa-Bitswap provides a significant improvement.

\section{Conclusion}\label{sec:conclusion}
In this paper we presented RaWa-Bitswap.
It improves privacy by prepending a privacy phase to the content discovery mechanism of Bitswap.
The evaluation of RaWa-Bitswap shows enhanced network-wide privacy with only a low performance overhead compared to Vanilla-Bitswap.
While our test setup with only 50 peers cannot capture all network dynamics, it shows promising results, even against active adversaries.

\section*{Acknowledgements}
We thank Protocol Labs for funding parts of our research.
Additionally, this work was supported by
the German Federal Ministry of Education and Research~(BMBF)
and the Saxon State Ministry for Science, Culture and Tourism~(SMWK)
by funding the competence center ScaDS.AI.

\IEEEtriggeratref{14}
\printbibliography[heading=bibintoc]

@INPROCEEDINGS{dingledine2004tor,
  AUTHOR = {Dingledine, Roger and Mathewson, Nick and Syverson, Paul F.},
  CROSSREF = {usenixsecurity04},
  DATE = {2004},
  PAGES = {303--320},
  TITLE = {Tor: The Second-Generation Onion Router},
}

@article{taylor2005gnutella,
  title={Gnutella},
  author={Taylor, Ian J},
  journal={From P2P to Web Services and Grids: Peers in a Client/Server World},
  pages={101--116},
  year={2005},
  publisher={Springer}
}

@article{wright2004predecessor,
  title={The predecessor attack: An analysis of a threat to anonymous communications systems},
  author={Wright, Matthew K and Adler, Micah and Levine, Brian Neil and Shields, Clay},
  journal={ACM Transactions on Information and System Security (TISSEC)},
  volume={7},
  number={4},
  pages={489--522},
  year={2004},
  publisher={ACM New York, NY, USA}
}

@ARTICLE{benet2014ipfs,
  TITLE = {{IPFS} - Content Addressed, Versioned, {P2P} File System},
  AUTHOR = {Benet, Juan},
  VOLUME = {abs/1407.3561},
  DATE = {2014-07},
  note = {http://arxiv.org/abs/1407.3561},
}

@INPROCEEDINGS{balduf2021monitoring,
  TITLE = {Monitoring Data Requests in Decentralized Data Storage Systems: A Case Study of IPFS},
  AUTHOR = {Balduf, Leonhard and Henningsen, Sebastian and Florian, Martin and Rust, Sebastian and Scheuermann, Bj{\"o}rn},
  PAGES = {658--668},
  CROSSREF = {icdcs22},
}

@article{de2021accelerating,
  TITLE = {Accelerating Content Routing with Bitswap: A Multi-Path File Transfer Protocol in IPFS and Filecoin},
  AUTHOR = {{De la Rocha}, Alfonso and Dias, David and Psaras, Yiannis},
  YEAR = {2021},
  PAGES = {11}
}

@inproceedings{trautwein2022design,
  title={Design and evaluation of IPFS: a storage layer for the decentralized web},
  author={Trautwein, Dennis and Raman, Aravindh and Tyson, Gareth and Castro, Ignacio and Scott, Will and Schubotz, Moritz and Gipp, Bela and Psaras, Yiannis},
  pages={739--752},
  CROSSREF={sigcomm22}
}

@INPROCEEDINGS{henningsen2020mapping,
  TITLE = {Mapping the Interplanetary Filesystem},
  AUTHOR = {Henningsen, Sebastian and Florian, Martin and Rust, Sebastian and Scheuermann, Bj{\"o}rn},
  PAGES = {289--297},
  CROSSREF = {networking2020}
}

@INPROCEEDINGS{shen2019understanding,
  TITLE = {Understanding {I/O} Performance of {IPFS} storage: a client's perspective},
  AUTHOR = {Shen, Jiajie and Li, Yi and Zhou, Yangfan and Wang, Xin},
  PAGES = {17:1--17:10},
  CROSSREF = {iwqos19},
}

@inproceedings{monteiro2022enriching,
  TITLE = {Enriching Kademlia by Partitioning},
  AUTHOR = {Monteiro, Joao and Costa, Pedro Akos and Leitao, Joao and De la Rocha, Alfonso and Psaras, Yiannis},
  PAGES = {33--38},
  CROSSREF = {icdcsw22},
}

@inproceedings{kanemitsu2023lookup,
  title={Lookup Parameter Optimization for Kademlia DHT Alternative in IPFS},
  author={Kanemitsu, Hidehiro and Kanai, Kenji and Nakazato, Hidenori},
  pages={905--913},
  crossref = {ipdpsw23},
}

@article{sridhar2023content,
  title={Content Censorship in the InterPlanetary File System},
  author={Sridhar, Srivatsan and Ascigil, Onur and Keizer, Navin and Genon, Fran{\c{c}}ois and Pierre, S{\'e}bastien and Psaras, Yiannis and Rivi{\`e}re, Etienne and Kr{\'o}l, Micha{\l}},
  journal={arXiv preprint arXiv:2307.12212},
  year={2023}
}

@article{bojja2017dandelion,
  title={Dandelion: Redesigning the Bitcoin network for anonymity},
  author={Bojja Venkatakrishnan, Shaileshh and Fanti, Giulia and Viswanath, Pramod},
  journal={Proceedings of the ACM on Measurement and Analysis of Computing Systems},
  volume={1},
  number={1},
  pages={1--34},
  year={2017},
  publisher={ACM New York, NY, USA}
}

@article{fanti2018dandelion++,
  title={Dandelion++ lightweight cryptocurrency networking with formal anonymity guarantees},
  author={Fanti, Giulia and Venkatakrishnan, Shaileshh Bojja and Bakshi, Surya and Denby, Bradley and Bhargava, Shruti and Miller, Andrew and Viswanath, Pramod},
  journal={Proceedings of the ACM on Measurement and Analysis of Computing Systems},
  volume={2},
  number={2},
  pages={1--35},
  year={2018},
  publisher={ACM New York, NY, USA}
}

@article{sharma2022anonymity,
  title={On the anonymity of peer-to-peer network anonymity schemes used by cryptocurrencies},
  author={Sharma, Piyush Kumar and Gosain, Devashish and Diaz, Claudia},
  journal={arXiv preprint arXiv:2201.11860},
  year={2022}
}

@INPROCEEDINGS{mclachlan2009scalable,
  TITLE = {Scalable Onion Routing with Torsk},
  AUTHOR = {McLachlan, Jon and Tran, Andrew and Hopper, Nicholas and Kim, Yongdae},  
  PAGES = {590--599},
  CROSSREF = {ccs09},
}

@inproceedings{freedman2002tarzan,
  title={Tarzan: A Peer-to-Peer Anonymizing Network Layer},
  author={Freedman, Michael J and Morris, Robert},
  pages={193--206},
  crossref = {ccs02},
}

@article{reiter1998crowds,
  title={Crowds: Anonymity for Web Transactions},
  author={Reiter, Michael K and Rubin, Aviel D},
  journal={{ACM} transactions on information and system security (TISSEC)},
  volume={1},
  number={1},
  pages={66--92},
  year={1998},
  publisher={ACM New York, NY, USA}
}

@article{franzoni2022clover,
  title={Clover: An Anonymous Transaction Relay Protocol for the Bitcoin P2P Network},
  author={Franzoni, Federico and Daza, Vanesa},
  journal={Peer-to-Peer Networking and Applications},
  volume = {15},
  number = {1},
  pages = {290--303},
  year = {2022},
}

@inproceedings{mittal2009shadowwalker,
  TITLE = {Shadowwalker: Peer-to-Peer Anonymous Communication using Redundant Structured Topologies},
  AUTHOR = {Mittal, Prateek and Borisov, Nikita},
  PAGES = {161--172},
  CROSSREF = {ccs09}
}

@article{mittal2012information,
  title = {Information Leaks in Structured Peer-to-Peer Anonymous Communication Systems},
  author = {Mittal, Prateek and Borisov, Nikita},
  journal = {{ACM} Transactions on Information and System Security (TISSEC)},
  volume = {15},
  number = {1},
  pages = {5:1--5:28},
  year={2012},
}

@INPROCEEDINGS{wang2010search,
  TITLE = {In Search of an Anonymous and Secure lookup: Attacks on Structured Peer-to-Peer Anonymous Communication Systems},
  AUTHOR = {Wang, Qiyan and Mittal, Prateek and Borisov, Nikita},
  PAGES = {308--318},
  CROSSREF = {ccs10}
}

@inproceedings{mislove2004ap3,
  title={{AP3}: Cooperative, decentralized anonymous communication},
  author = {Mislove, Alan and Oberoi, Gaurav and Post, Ansley and Reis, Charles and Druschel, Peter and Wallach, Dan S},
  pages={30},
  crossref = {sigops04},
}

@inproceedings{qian2016garlic,
  title={Garlic Cast: Lightweight and Decentralized Anonymous Content Sharing},
  author={Qian, Chen and Shi, Junjie and Yu, Zihao and Yu, Ye and Zhong, Sheng},
  pages={216--223},
  crossref = {icpads16},
}

@article{liu2011rumor,
  title={Rumor Riding: Anonymizing Unstructured Peer-to-Peer Systems},
  author={Liu, Yunhao and Han, Jinsong and Wang, Jilong},
  journal={{IEEE} Transactions on Parallel and Distributed Systems},
  volume={22},
  number={3},
  pages={464--475},
  year={2011},
}

@article{daniel2024exploring,
  title={Exploring the design space of privacy-enhanced content discovery for bitswap},
  author={Daniel, Erik and Tschorsch, Florian},
  journal={Computer Communications},
  volume={217},
  pages={12--24},
  year={2024},
}

@inproceedings{daniel2023improving,
  title={Improving Bitswap Privacy with Forwarding and Source Obfuscation},
  author={Daniel, Erik and Ebert, Marcel and Tschorsch, Florian},
  pages={1--4},
  crossref = {lcn23},
}

@inproceedings{daniel2022passively,
  TITLE={Passively Measuring IPFS Churn and Network Size},
  AUTHOR={Daniel, Erik and Tschorsch, Florian},
  PAGES={60--65},
  CROSSREF={icdcsw22},
}

@MISC{gitgoboxo,
  TITLE = {Github -- ipfs/boxo: A set of reference libraries for building IPFS applications and implementations in Go.},
  AUTHOR = {{Protocol Labs}},
  PUBLISHER = {GitHub},
  JOURNAL = {GitHub repository},
  howpublished={\url{https://github.com/ipfs/boxo}},
  note={Accessed: 2023-12}
}

@PROCEEDINGS{ccs02,
  BOKTITLE = {CCS~'02: Proceedings of the 9th {ACM} Conference on Computer and Communications Security},
  LOCATION = {Washington, DC, USA},
  DATE = {2002-11},
}

@PROCEEDINGS{sigops04,
  BOOKTITLE = {SIGOPS~'04: Proceedings of the 11st {ACM} {SIGOPS} European Workshop},
  LOCATION = {Leuven, Belgium},
  DATE = {2004-09},
}

@PROCEEDINGS{usenixsecurity04,
  LOCATION = {San Diego, CA, USA},
  BOOKTITLE = {USENIX Security~'04: Proceedings of the 13th USENIX Security Symposium},
  DATE = {2004-08},
}

@PROCEEDINGS{ccs09,
  BOOKTITLE = {CCS~'09: Proceedings of the 16th {ACM} Conference on Computer and Communications Security},
  LOCATION = {Chicago, IL, USA},
  DATE = {2009-11}
}

@PROCEEDINGS{ccs10,
  BOOKTITLE={CCS~'10: Proceedings of the 17th {ACM} Conference on Computer and Communications Security},
  LOCATION = {Chicago, IL, USA},
  DATE = {2010-10}
}

@proceedings{icpads16,
  BOOKTITLE = {ICPADS~'16: Proceedings of the 22nd {IEEE} International Conference on Parallel and Distributed Systems},
  LOCATION = {Wuhan, China},
  DATE = {2016-12},
}

@PROCEEDINGS{iwqos19,
  LOCATION = {Phoenix, AZ, USA},
  BOOKTITLE = {IWQoS~'19: Proceedings of the International Symposium on Quality of Service},
  DATE = {2019-06}
}

@PROCEEDINGS{networking2020,
  LOCATION = {Paris, France},
  BOOKTITLE = {Networking~'20: Proceedings of the 19th IFIP Networking Conference},
  DATE = {2020-06}
}

@Proceedings{icdcs22,
  BOOKTITLE = {ICDCS~'22: Proceedings of the 42nd {IEEE} International Conference on Distributed Computing Systems},
  LOCATION = {Bologna, Italy},
  DATE = {2022-07},
}

@Proceedings{icdcsw22,
  BOOKTITLE = {ICDCSW~'22: Proceedings of the 42nd {IEEE} International Conference on Distributed Computing Systems Workshops},
  LOCATION = {Bologna, Italy},
  DATE = {2022-07},
}

@Proceedings{sigcomm22,
  BOOKTITLE = {SIGCOMM'~22: Proceedings of the {ACM} {SIGCOMM} 2022 Conference},
  LOCATION = {Amsterdam, The Netherlands},
  DATE = {2022-08},
}

@PROCEEDINGS{ipdpsw23,
  BOOKTITLE = {IPDPSW~'23: Proceedings of the {IEEE} International Parallel and Distributed Processing Symposium - Workshops},
  LOCATION = {St. Petersburg, FL, USA},
  DATE = {2023-05},
}

@PROCEEDINGS{lcn23,
  BOOKTITLE = {LCN'~23: In the Proceedings of the {IEEE} 48th Conference on Local Computer Networks},
  LOCATION = {Daytona Beach, FL, USA},
  DATE = {2023-10},
}

\end{document}